\documentclass[runningheads]{svmult}

\usepackage{mathptmx}       
\usepackage{helvet}         
\usepackage{courier}        
\usepackage{type1cm}        
\usepackage{makeidx}   
\usepackage{graphicx}  
\usepackage{multicol}  
\usepackage[bottom]{footmisc}
\makeindex             

\begin{document}
\title*{Prospects for Studies of Stellar Evolution and Stellar Death 
in the {\em JWST} Era}
\toctitle{Stellar Evolution and Stellar Death - the Potential of the JWST Era}
%
%
\titlerunning{Stellar Evolution and Stellar Death}
%
\author{Michael J. Barlow}

\authorrunning{M. J. Barlow}
%
%
\institute{Department of Physics \& Astronomy, \\
           University College London, \\
           Gower Street, London WC1E 6BT, U.K.}

\maketitle              

\begin{abstract}
I review the prospects for studies of the advanced evolutionary stages of 
low-, intermediate- and high-mass stars by the {\em JWST} and concurrent 
facilities, with particular emphasis on how they may help elucidate the 
dominant contributors to the interstellar dust component of galaxies. 
Observations extending from the mid-infrared to the submillimeter can help 
quantify the heavy element and dust species inputs to galaxies from AGB 
stars. {\em JWST}'s MIRI mid-infrared instrument will be so sensitive that 
observations of the dust emission from individual intergalactic AGB stars 
and planetary nebulae in the Virgo Cluster will be feasible. The {\em 
Herschel Space Observatory} will enable the last largely unexplored 
spectral region, from the far-IR to the submm, to be surveyed for new 
lines and dust features, while {\em SOFIA} will cover the 
wavelength gap between {\em JWST} and {\em Herschel}, a spectral region 
containing important fine structure lines, together with key water-ice
and crystalline silicate bands. {\em Spitzer} has significantly increased
the number of Type~II supernovae that have been surveyed for early-epoch
dust formation but reliable quantification of the dust contributions
from massive star supernovae of  Type~II, Type~Ib and Type~Ic 
to low- and high-redshift galaxies should come from {\em JWST} MIRI
observations, which will be able to probe a volume over 1000 times
larger than {\em Spitzer}. 
\end{abstract}

\section{Introduction}

The bulk of the heavy element enrichment of the interstellar media of
galaxies is a result of mass loss during the final evolutionary stages of
stars, yet these final stages are currently the least well understood
parts of their lives. This is because (a) relatively short evolutionary
timescales often lead to comparatively small numbers of objects in
different advanced evolutionary stages; (b) strong mass loss with dust
formation during some late phases can lead to self-obscuration at optical
wavelengths, requiring infrared surveys for their detection; (c) the mass
loss process itself can have a complex dependence on pulsational
properties, metallicity and the physics of dust formation. For low and
intermediate mass stars ($\leq$ 8~M$_{\odot}$), strong mass loss during
the asymptotic giant branch (AGB) phase (Iben \& Renzini 1983) exposes the
dredged-up products of nucleosynthesis (nitrogen from CNO-cycle H-shell
burning and carbon from helium-shell burning). For higher mass stars, mass
loss stripping eventually exposes the products of H- and He-burning.
During the late stages of massive stars these products can dominate their
spectra, as luminous blue variables (LBVs) or as WN, WC or WO Wolf-Rayet
(WR) stars.

Current or planned optical and infrared surveys of the Milky Way Galaxy
(MWG) and of other Local Group galaxies will yield much larger samples of
single and binary stars in the various brief evolutionary phases that
occur near the end-points of their lives. Ground-based optical surveys
include the nearly complete INT Photometric H$\alpha$ Survey of the
northern galactic plane (IPHAS; Drew et al. 2005) and the complementary
VST Photometric H$\alpha$ Survey (VPHAS+) of the southern galactic plane
that will begin in 2009, while ground-based near-IR surveys include the
UKIRT Infrared Deep Sky Survey (UKIDSS; Lawrence et al. 2007) in the
north, and the VISTA sky surveys in the south. From space, we have the
recent {\em Spitzer} GLIMPSE (Benjamin et al. 2003) and MIPSGAL (Carey 
et al. 2005) mid-IR surveys of the MWG between
$l = \pm60^{\rm o}$, which will be joined by the {\em Herschel}
70-500~$\mu$m Hi-GAL Survey of the same MWG regions, while the {\em
Spitzer} SAGE survey (Meixner et al. 2006) and the {\em Herschel} Heritage
survey will map the contents of the lower metallicity LMC and SMC dwarf
galaxies from 3.6 to 500~$\mu$m. These surveys will enable many stars in
different advanced evolutionary stages to be identified, from their
characteristic optical/IR spectral energy distributions (SEDs).

The {\em JWST} is an infrared observatory and a significant number of its
concurrent facilities will operate at infrared and submillimeter
wavelengths. The gaseous component of mass loss outflows can be studied in
detail at these wavelengths using high resolution spectrometers. In
addition, towards the end-points of their evolution many stars become
extremely luminous at infrared wavelengths due to re-emission by dust
particles formed in their outflows. This dust emission can be used as a
tool for studying stellar populations, for studying the mass loss process
and its history and for investigating the effects of stellar evolution on
the enrichment of galaxies. Carbon and silicate dust particles are thought
to be ubiquitous throughout our Galaxy but where are these particles
formed? Evolved stars (AGB stars, M supergiants, WR stars, supernovae,
etc.) are known to be significant sources of dust particles but what are
the dominant sources of dust in our own and other galaxies? Tielens,
Waters and Bernatowicz (2005) summarised current estimates for the gas and
dust inputs to the ISM of our galaxy from various classes of evolved
stars; their review pinpoints the fact that the integrated contributions
by some of these stellar types have large uncertainties at present.

As a result of ongoing measurements of mass loss rates by the {\em
Spitzer} SAGE Survey, the dominant stellar dust contributors to the
metal-poor LMC and SMC dwarf galaxies, at known distances, will soon be
known. For the much more massive Milky Way Galaxy, the optical and IR
surveys discussed above will lead to much larger samples of hitherto rare
evolved star types. Once accurate parallaxes become available for one
billion MWG stars down to 20th magnitude from ESA's GAIA Mission 
(2012-20),
then we will have reliable distances, population numbers and mass loss
rates for a wide range of evolutionary types, allowing accurate gas and
dust enrichment rates for the MWG to be established. In this review I will
focus on future observations by the {\em JWST} and other new facilities of
low-, intermediate- and high-mass stars in the advanced evolutionary
phases that are likely to provide the dominant contributions to the gas 
and dust enrichment of galaxies.

\section{AGB stars, post-AGB objects and planetary nebulae}

\subsection{The infrared spectra of evolved stars}
Figure~1 illustrates the mid-infrared spectra of eight oxygen-rich stars 
that are at different stages in their evolution up the AGB (Sylvester et 
al. 1999). 

\begin{figure}
\begin{center}
\includegraphics[width=0.65\textwidth]{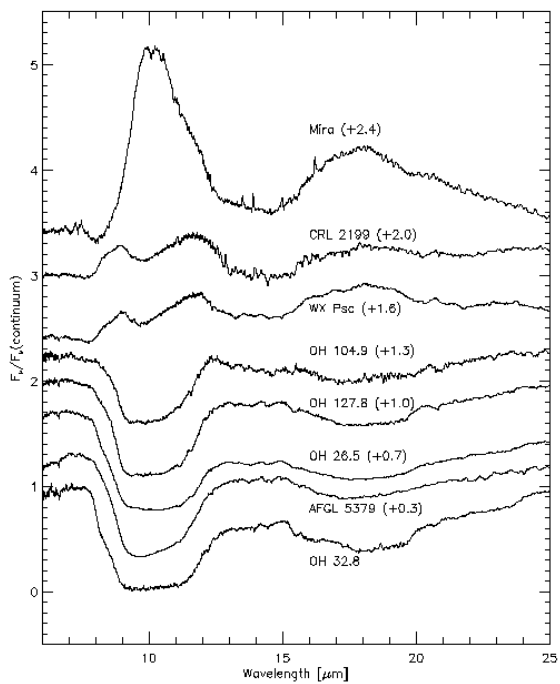}
\end{center}

\caption[]{The 6-25-$\mu$m {\em ISO} SWS spectra of eight oxygen-rich AGB 
stars,
in order of increasing mass loss rate, from top to bottom. The
broad amorphous silicate features at 9.7~$\mu$m and 18~$\mu$m are 
increasingly self-absorbed with increasing mass loss rate (from Sylvester 
et al. 1999).}
\end{figure}

Mira (Omicron 
Ceti) exhibits pure emission silicate bands at 10 and 18~$\mu$m, while 
CRL~2199 and WX~Psc (=IRC+10011) show self-absorbed emission features and 
the remaining five sources show absorption features. This sequence is 
interpreted as one of increasing mass loss rate and increasing 
circumstellar self-absorption. Although the silicate-absorption objects 
are more luminous, the time spent at such high luminosities and mass loss 
rates is much shorter than the time spent lower down the AGB.
The mid-IR spectra of ellliptical galaxies show 10-$\mu$m silicate 
features in emission (Bregman, Temi \& Bregman 2006), interpreted as 
resulting from the combined spectra of many AGB stars having silicate 
emission features.

\begin{figure}
\begin{center}
\includegraphics[width=0.6\textwidth]{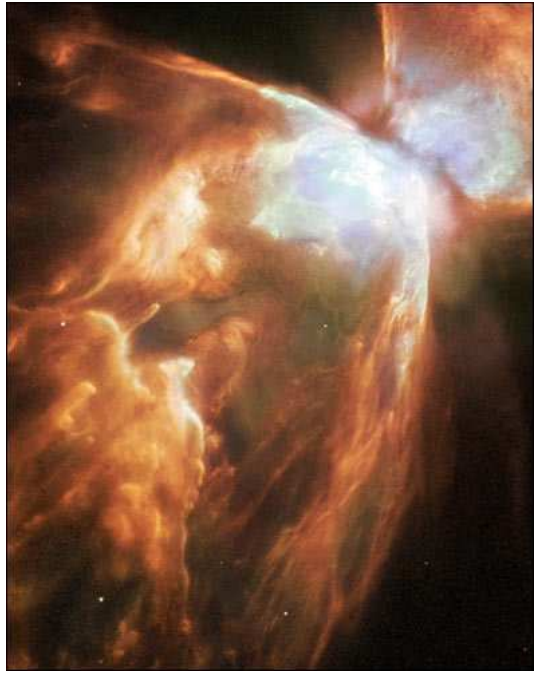}
\end{center}

\caption[]{{\em HST} WFPC2 F656N image of the central parts of NGC 6302,
showing the obscured region (top right) attributed to an edge-on
dust disk that is collimating the bipolar
outflow geometry. See Matsuura et al. (2005).}
\end{figure}

For the objects shown in Fig.~1, Sylvester et al. (1999) found that after
division by a smooth continuum fit, a number of sharp features were
evident longwards of 20~$\mu$m that corresponded very well to crystalline
silicate emission features that had previously been discovered in the {\em
ISO} spectra of several classes of objects, including comets, very young
stars and very evolved objects. An example of the latter is NGC~6302, a
likely descendant of an extreme OH/IR star. It has a massive edge-on dust
disk which completely obscures the central star, whose effective
temperature is estimated from photoionization modelling to exceed
200,000~K. An {\em HST} image of NGC~6302 is shown in Figure~2. Its {\em
ISO} spectrum (Figures 3 and 4) exhibits many sharp crystalline silicate
emission features, a well as features due to other species, such as
crystalline water ice.

\begin{figure}
\begin{center}
\includegraphics[width=0.67\textwidth]{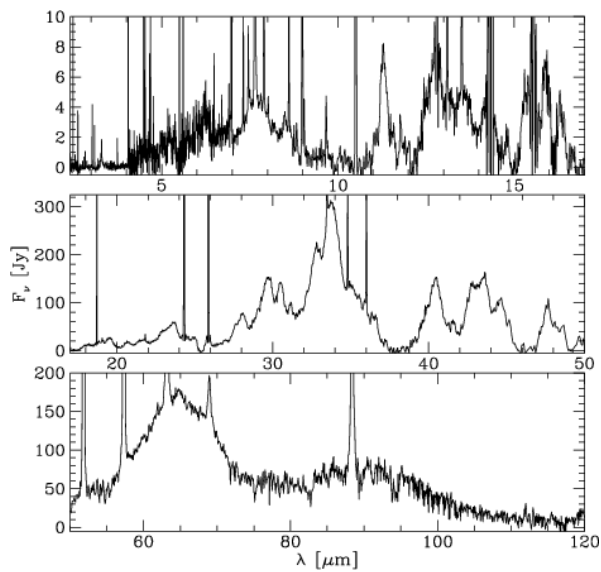}
\end{center}

\caption[]{The continuum-subtracted {\em ISO} SWS+LWS spectrum of the 
extreme 
bipolar planetary nebula NGC~6302, which has a massive edge-on dust disk, 
illustrating the presence of PAH features (top panel), as well as many 
crystalline silicate and water-ice features (bottom two panels).
From Molster et al. (2001).} 
\end{figure}

\begin{figure}
\begin{center}
\includegraphics[width=0.73\textwidth]{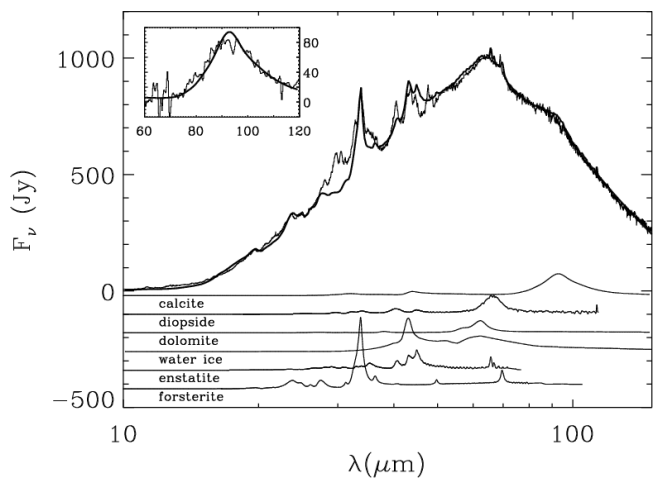}
\end{center}

\caption[]{A multi-component continuum and dust feature fit to the {\em 
ISO} far-infrared spectrum of 
NGC~6302 (Kemper et al. 2002). {\em JWST}-MIRI stops at 28~$\mu$m. {\em 
Herschel}-PACS starts at 57~$\mu$m. In between, there are many crystalline 
silicate bands, plus the crucial 44-$\mu$m crystalline ice band and
the [O~{\sc iii}] 52-$\mu$m line. In the next ten years, only {\em SOFIA} 
will be able to observe the 28-57-$\mu$m wavelength range.}
\end{figure}

The crystalline particles in NGC~6302 and other sources are thought to
have been produced by the annealing of amorphous grains following upward
temperature excursions. The peak wavelengths and widths of some of the
features have been found to be good indicators of the temperature of the
emitting particles. Figure 5 (left panel) shows laboratory spectra of the
69-$\mu$m band of forsterite (crystalline olivine), obtained at three
different temperatures (Bowey et al. 2002). As the temperature changes
from 295~K to 3.5~K, the peak wavelength of the band moves shortwards by
nearly a micron, while the FWHM of the feature becomes narrower. The
temperature derived from the peak wavelength of the 69-$\mu$m forsterite
band (Fig.~5; right panel) is well-correlated with the continuum dust
temperature and thus is an excellent dust thermometer that can be used by
{\em Herschel}-PACS and by {\em SOFIA} for studies of post-MS and pre-MS
objects.

\begin{figure}
\begin{center}
\includegraphics[width=0.35\textwidth]{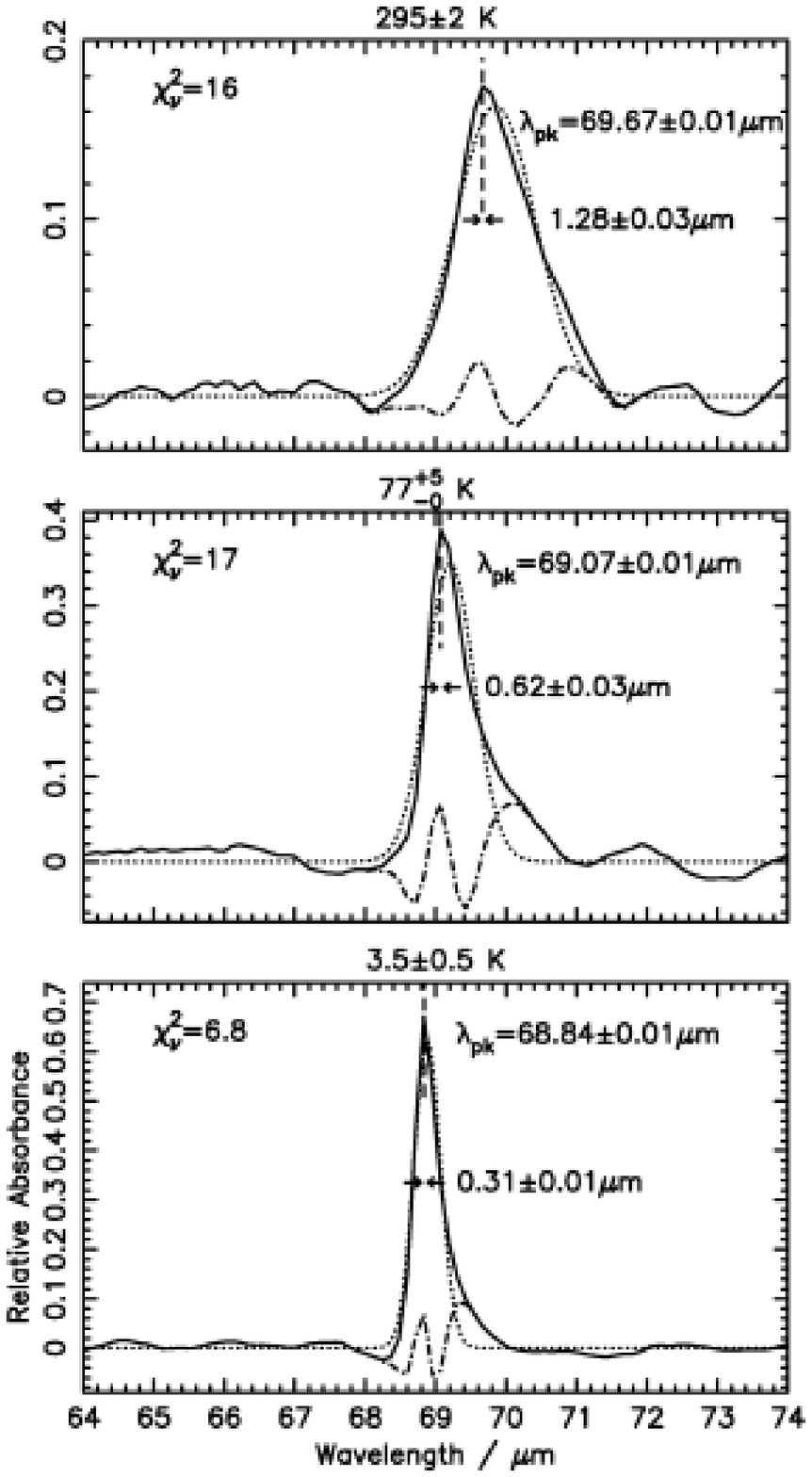}
\includegraphics[width=0.60\textwidth]{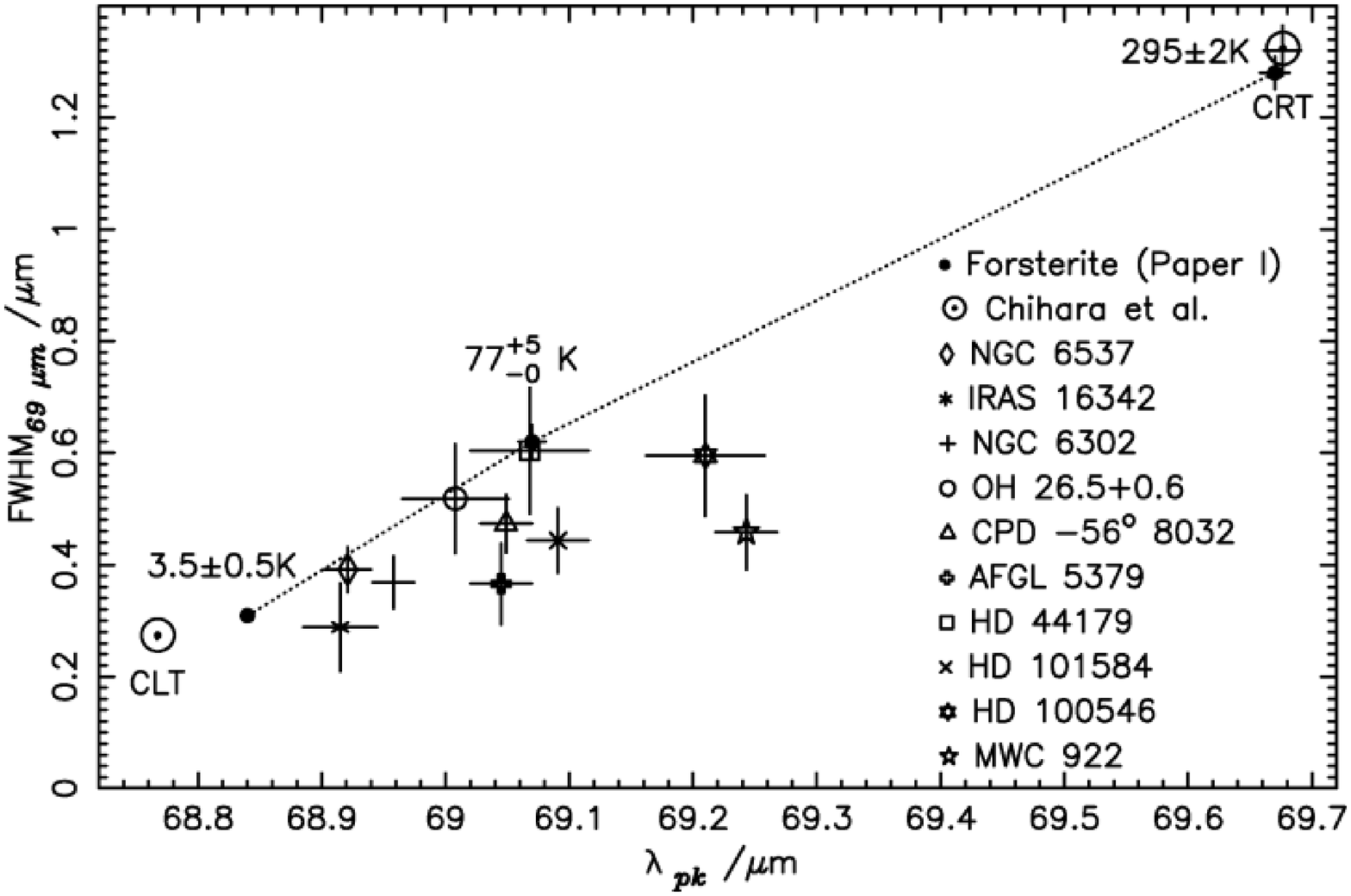}
\end{center}

\caption[]{Left: Laboratory measurements of the 69-$\mu$m forsterite band 
FWHM vs. peak wavelength, for three different temperatures Right: The FWHM 
versus observed peak wavelength of the 69-$\mu$m feature for ten 
astronomical sources (Bowey et al. 2002).}
\end{figure}

\begin{figure}
\begin{center}
\includegraphics[width=0.7\textwidth]{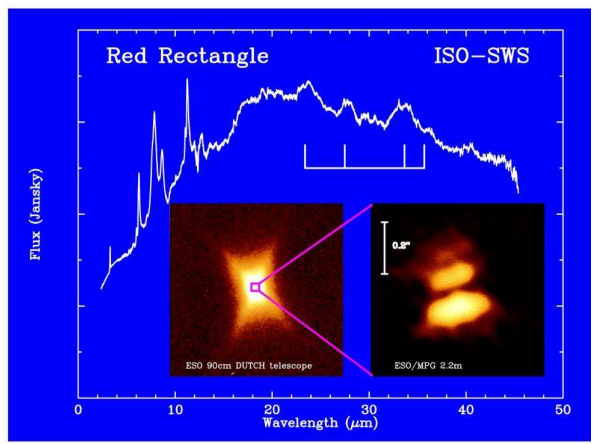}
\end{center}

\caption[]{The {\em ISO} SWS 2.4-45-$\mu$m spectrum of HD~44179, the Red 
Rectangle, illustrating its strong PAH emission features shortwards of
15~$\mu$m and its prominent crystalline silicate features longwards
of 20~$\mu$m. The insets show (left) a ground-based optical image of the 
Red Rectangle, and (right) an interferometric optical image
of the central region, showing a dark dust lane cutting across
the center. Figure courtesy of Frank Molster and Rens Waters}
\end{figure}

As well as exhibiting strong crystalline silicate emission features 
longwards of 20~$\mu$m, NGC~6302 also exhibits a number of emission bands 
in the 5-15~$\mu$m region of its spectrum that are usually attributed to 
C-rich PAH particles (Figure~3, panel 1). This `dual dust chemistry' 
phenomenon has also been encountered in the infrared spectra of a number 
of other post-AGB objects and planetary nebulae, particularly objects 
having late WC-type central stars (Cohen et al. 2002). An archetypal 
example of the dual dust chemistry phenomenon is the Red Rectangle bipolar 
nebula around the post-AGB star HD~44179 (Figure~6). Longwards of 
20~$\mu$m, its spectrum is dominated by strong crystalline silicate 
emission features, whereas shortwards of 15~$\mu$m it is dominated by 
strong PAH emission bands. The current most favoured interpretation
of dual dust chemistry sources is that the crystalline silicates
reside in a cool shielded dust disk around a binary system, having
been captured there following an earlier phase of AGB mass loss; later
on, the AGB star evolved from an O-rich to a C-rich chemistry, following 
several episodes of the 3rd dredge-up, with newly produced
C-rich particles being channeled into polar outflows by the O-rich
dust disk.

\begin{figure}
\begin{center}
\includegraphics[width=0.50\textwidth]{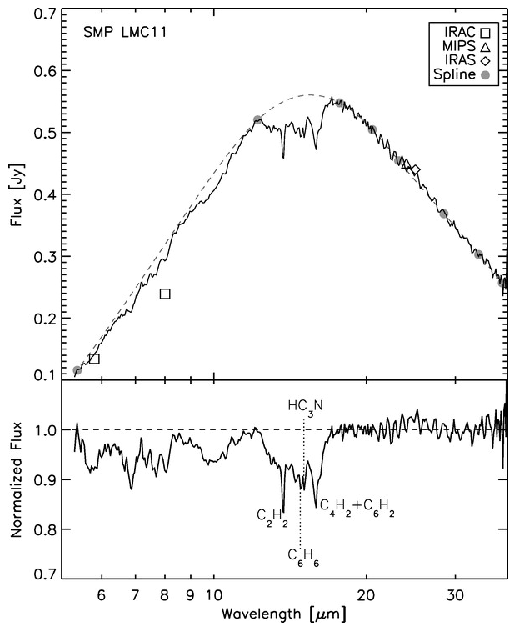}
\includegraphics[width=0.45\textwidth, bb=190 230 9 9]{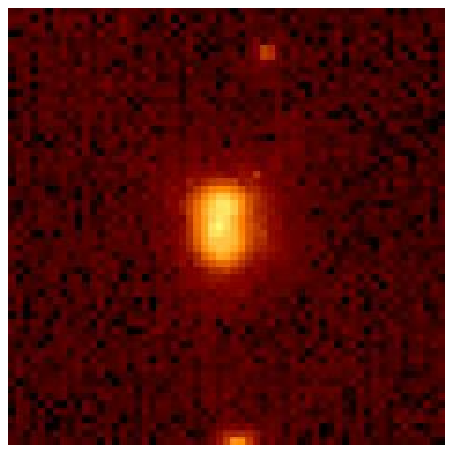}
\includegraphics[width=0.85\textwidth]{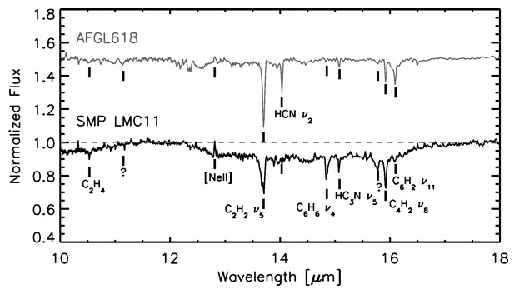}
q\end{center}

\caption[]{Top-left: The {\em Spitzer}-IRS spectrum of the 
proto-planetary nebula SMP LMC11 (Bernard-Salas et al. 2006)
showing both fluxed and normalized versions, with several hydrocarbon 
absorption features identified. Bottom: Normalized 10-18-$\mu$m
spectra of AFGL~618 (Cernicharo et al. 2001) and SMP LMC11
(Bernard-Salas et al. 2006), showing absorption features due to
benzene and other complex hydrocarbons. Top-right: {\em HST} V-band
image of SMP LMC11, from Shaw et al. (2006), showing its bipolar
structure.}

\end{figure}

AFGL~618 and AFGL~2688 are Galactic carbon-rich bipolar post-AGB objects
which, unlike the Red Rectangle, do not exhibit dual dust chemistries,
with only carbon-rich molecular or dust features having been detected in
their spectra. Cernicharo et al. (2001) identified absorption features due
to benzene and several other complex hydrocarbon molecules in the {\em
ISO} SWS spectrum of AFGL~618, while Bernard-Salas et al. (2006) have
detected the same absorption features in an R~=~600 {\em Spitzer} IRS
spectrum of SMP~LMC11 (Figure~7), an LMC object that has been classified
in the past as a planetary nebula, but whose IR spectrum and {\em HST} 
image
(Fig.~7) indicate it as likely to be transiting between the post-AGB and
ionized PN phases. As can be seen in Fig.~7, the 14.85-$\mu$m benzene
absorption feature is even stronger in the spectrum of SMP~LMC11 than in
the spectrum of AFGL~618.

\subsection{Potential JWST studies of extragalactic evolved stars and 
planetary nebulae}

SMP~LMC11 is an example of a potential target for higher spectral
resolution infrared spectroscopy by {\em SOFIA} (apart from the
13.5-16.5~$\mu$m region, which is inaccessible even from airborne
altitudes) or by {\em JWST}-MIRI. However, it is worth bearing in mind
that even at the distance of the LMC, SMP~LMC11, with a peak flux of
$\sim$0.5~Jy (Fig.~7), will only just be observable using MIRI's R~=~3000
integral field spectroscopy mode, given the expected 0.5~Jy saturation
limit of this mode. This starkly illustrates the huge sensitivity gains
that MIRI will confer relative to previous mid-infrared instrumentation.
Photometric studies with MIRI of AGB stars and post-AGB nebulae will
generally only be feasible for targets lying at much greater distances
than the Magellanic Clouds.

Using {\em JWST}'s MIRI in its $1.3' \times 1.7'$ imaging mode, SMP~LMC11
would be detectable at 10$\sigma$ in 10$^4$~sec out to a distance of
18~Mpc at 18~$\mu$m; out to a distance of 38~Mpc at 10~$\mu$m; and out to
a distance of 47~Mpc at 7.5~$\mu$m. Objects discovered via imaging
photometry could be followed up from 5-10~$\mu$m with MIRI's long-slit
R~=~100 spectrometer. SMP~LMC11 would yield 10$\sigma$ per spectral
resolution element in 10$^4$~sec at a distance of 14.5~Mpc. These numbers
indicate that in the above modes MIRI will be comfortably capable of
studying similar objects out to the Virgo Cluster (D $\sim$ 14~Mpc) and
beyond.

\subsubsection{A case study: the intergalactic stellar population of the 
Virgo Cluster}

In recent years planetary nebulae have become important probes of
extragalactic stellar systems. Up to 10\% of the total luminosity of a
planetary nebula, $\sim$500~L$_{\odot}$, can be emitted in the dominant
cooling line, [O~{\sc iii}] $\lambda$5007. This, coupled with the
narrowness of the line ($\sim$15-25~km~s$^{-1}$), makes it extremely easy
to detect PNe in external galaxies using a narrow-band filter tuned to the
galaxy redshift.

Arnaboldi et al. (1996) measured velocities for 19 PNe in the outer
regions of the giant elliptical galaxy NGC~4406, in the southern Virgo
extension region. Although this galaxy has a peculiar radial velocity of
$-227$~km~s$^{-1}$, three of the PNe had velocities close to 
1400~km~s$^{-1}$,
the mean radial velocity of the Virgo cluster. It was concluded that they
were intracluster PNe. Theuns \& Warren (1997) discovered ten PN
candidates in the Fornax cluster, in fields well away from any Fornax
galaxy - consistent with tidal stripping of cluster galaxies. They
estimated that intracluster stars could account for up to 40\% of all the
stars in the Fornax cluster. Mendez et al. (1997) surveyed a 50~arcmin$^2$
area near the centre of the Virgo cluster, detecting 11 PN candidates.
From this, they estimated a total stellar mass of about 
4$\times10^9$~M$_\odot$ in 
their survey area and that such a population could account for up to 50\% 
of the total stellar mass in the Virgo cluster.

Follow-up observations have confirmed large numbers of intergalactic PN
candidates in Virgo. However, they represent a tiny tip of a very large
iceberg. How to sample more of the iceberg? The discussion above of {\em
JWST} MIRI capabilities indicates that mid-infrared observations
of dust-emitting AGB stars and planetary nebulae located in
intergalactic regions of the Virgo Cluster are feasible.

\begin{figure}
\begin{center}
\includegraphics[width=0.49\textwidth]{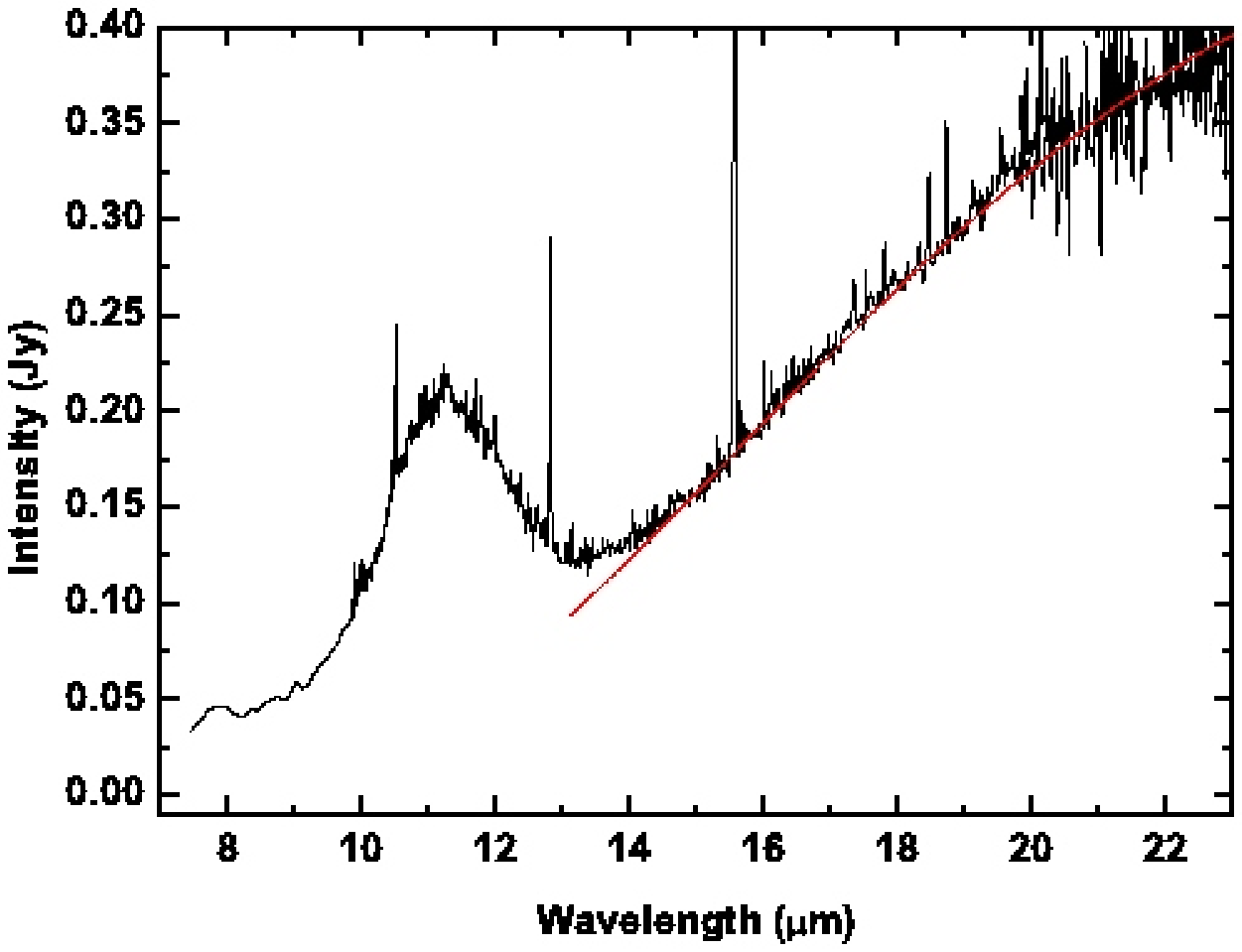}
\includegraphics[width=0.47\textwidth]{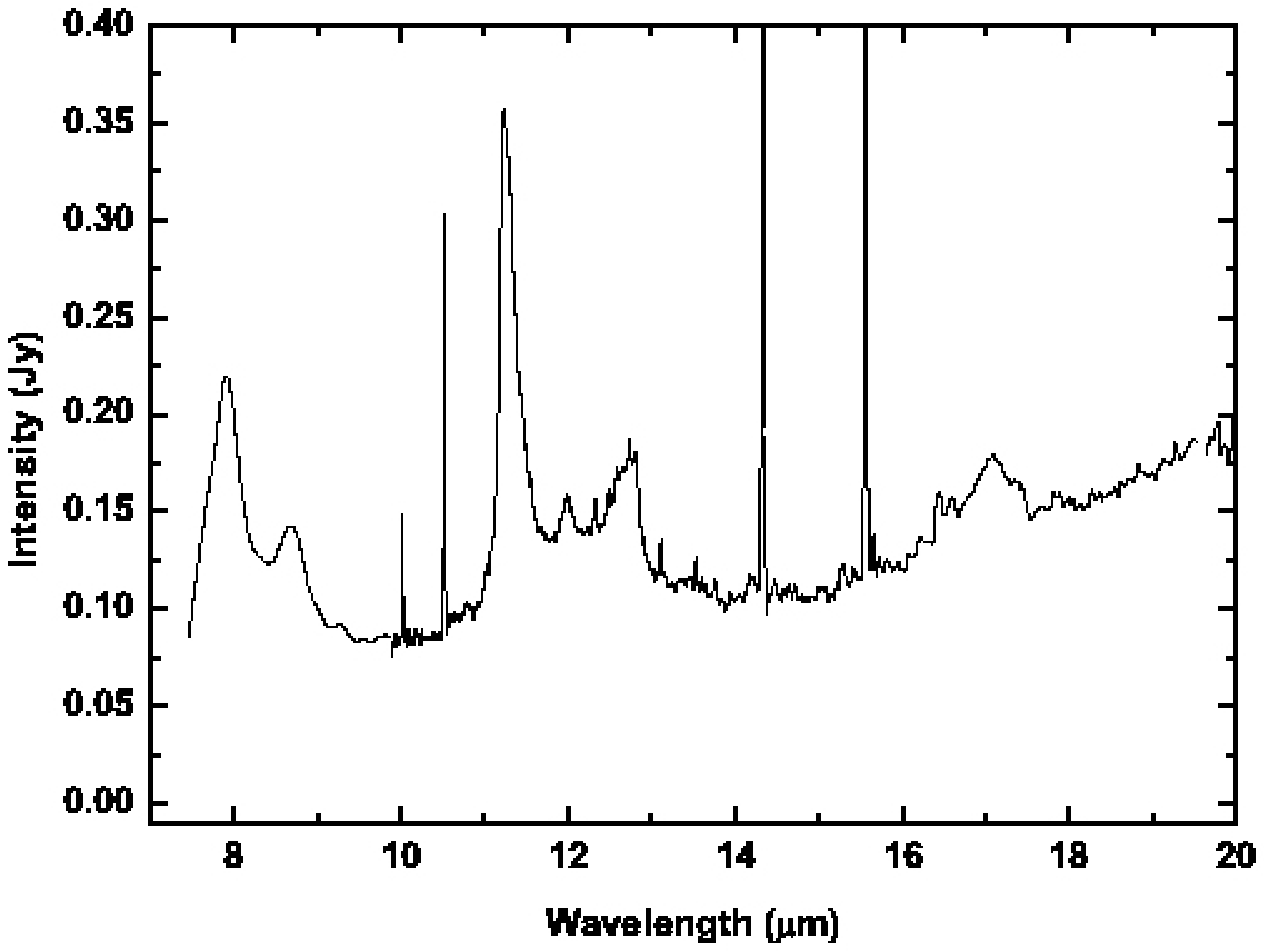}
\end{center}

\caption[]{{\em Spitzer}-IRS spectra of (left) SMP LMC8, showing a strong 
SiC 11.2-$\mu$m emission feature and (right) SMP LMC36, showing very 
strong PAH emission bands. Both planetary nebulae also exhibit narrow
fine structure ionic emission lines in their spectra.} 

\end{figure}

\begin{figure}
\begin{center}
\includegraphics[width=0.47\textwidth]{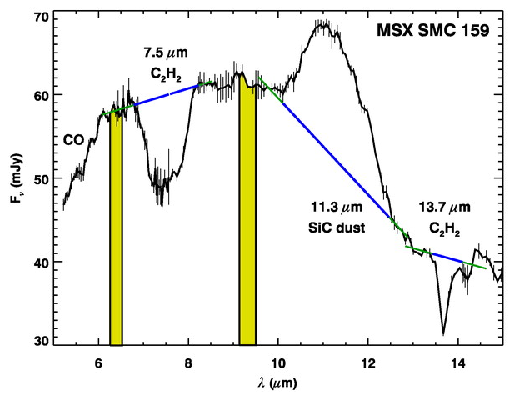}
\includegraphics[width=0.51\textwidth]{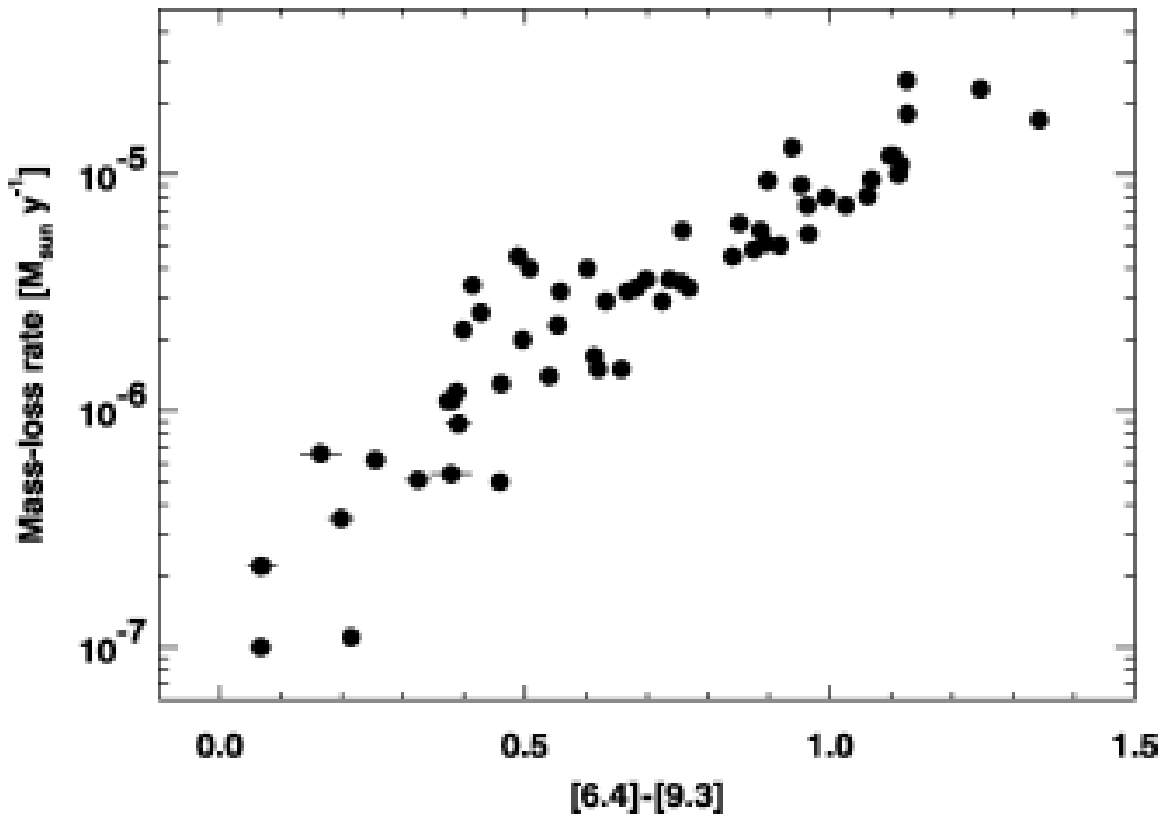}
\end{center}

\caption[]{Left: the {\em Spitzer} IRS spectrum of the carbon star MSX 
SMC~159,
illustrating the 11.2-$\mu$m SiC dust emission feature, as well as several 
important molecular absorption features and the narrow `continuum' regions
defined at 6.4~$\mu$m and 9.3~$\mu$m by Sloan et al. (2006). Right: a plot
from Groenewegen et al. (2007) of the correlation between the {\em 
Spitzer} [6.4]-[9.3] colors of SMC carbon stars and their derived mass 
loss rates.
}
\end{figure}

Figure~8 shows the {\em Spitzer} mid-infrared spectra of two LMC planetary
nebulae. The fluxes measured for the SiC-emitting nebula SMP~LMC8
indicate that in 10$^4$~sec MIRI imaging would yield 10$\sigma$ detections
for a distance of D~=~19~Mpc with the 7.7- and 11.3-$\mu$m filters and for
a distance of 13~Mpc with the 18-$\mu$m filter. For the strongly
PAH-emitting PN SMP~LMC36, in 10$^4$~sec MIRI imaging would yield a
10$\sigma$ detection for a distance of 44~Mpc with the 7.7-$\mu$m filter
and for a distance of 23~Mpc with the 11.3-$\mu$m filter.
The carbon star MSX~SMC~159, whose {\em Spitzer} IRS spectrum is shown in
Figure~9 (from Sloan et al. 2006), would yield 10$\sigma$ detections in
10$^4$~sec with MIRI out to a distance of D~=29~Mpc with the 5.6-$\mu$m
filter, to 23~Mpc with the 7.7-$\mu$m filter and to 12~Mpc with the
11.3-$\mu$m filter. Similar detection limits apply for oxygen-rich AGB
stars. As shown by Groenewegen et al. (2007; see Fig.~9), mid-infrared
colour indices can be used to estimate mass loss rates and thereby mass
inputs from stellar populations hosting AGB stars.

As well as MIRI, the other {\em JWST} instruments will also be useful for
such studies, e.g. with the JWST Tunable Filter Imager (R~=~100), a carbon
star similar to MSX~SMC~159 would give 10$\sigma$ per resolution element
in 10$^4$~sec for distances out to D~=~24-29~Mpc, for wavelengths from
$2-4~\mu$m. There are at least 20-30 AGB stars for every PN, so there
should be many detectable AGB stars in Virgo Cluster fields, allowing the
total stellar population to be determined, as well as their gas and dust
mass inputs into the intracluster medium.

\subsection{Exploring new wavelength regions at high spectral resolution}

The 200-650~$\mu$m range is the last largely unexplored region of the
astronomical spectrum. Many spectral lines and features were observed for
the first time in the 45-200~$\mu$m region by {\em ISO}'s Long Wavelength
Spectrometer, e.g. Figure~10 shows the LWS spectra of the oxygen-rich M
supergiant VY~CMa and the carbon star IRC+10$^{\rm o}$216. Similar numbers
of new lines and features can be expected to be found in the
200-650~$\mu$m spectral region. The HIFI, PACS and SPIRE spectrometers
onboard ESA's {\em Herschel Space Observatory} will observe a large number
of targets in this wavelength range. HIFI is capable of obtaining complete
spectral scans from 157-625~$\mu$m at a resolving power of R~=~10$^6$ for
a range of archetypal O-rich and C-rich sources, which
would allow an unprecedented line inventory to be built up. By measuring
line fluxes and profiles with high spectral resolution, HIFI and later
ALMA will be able to probe the dynamics of stellar wind outflows and to
use a wide range of atomic and molecular species to study the wind
chemistry and thermal structure as a function of distance from the central
source, with the strong cooling lines of CO, HCN and H$_2$O being
particularly
well-suited to probing the thermal and density structures of stellar
winds. The superb sensitivity of ALMA will allow detailed
studies to be made of objects located throughout the Milky Way and in the 
Magellanic Clouds. In addition, its high angular resolution will 
make it very well-suited to probe the dynamical and physical conditions
in the complex nebulae found around some AGB stars, post-AGB objects and
planetary nebulae.

\begin{figure}
\includegraphics[width=0.65\textwidth, bb=0 25 300 290]{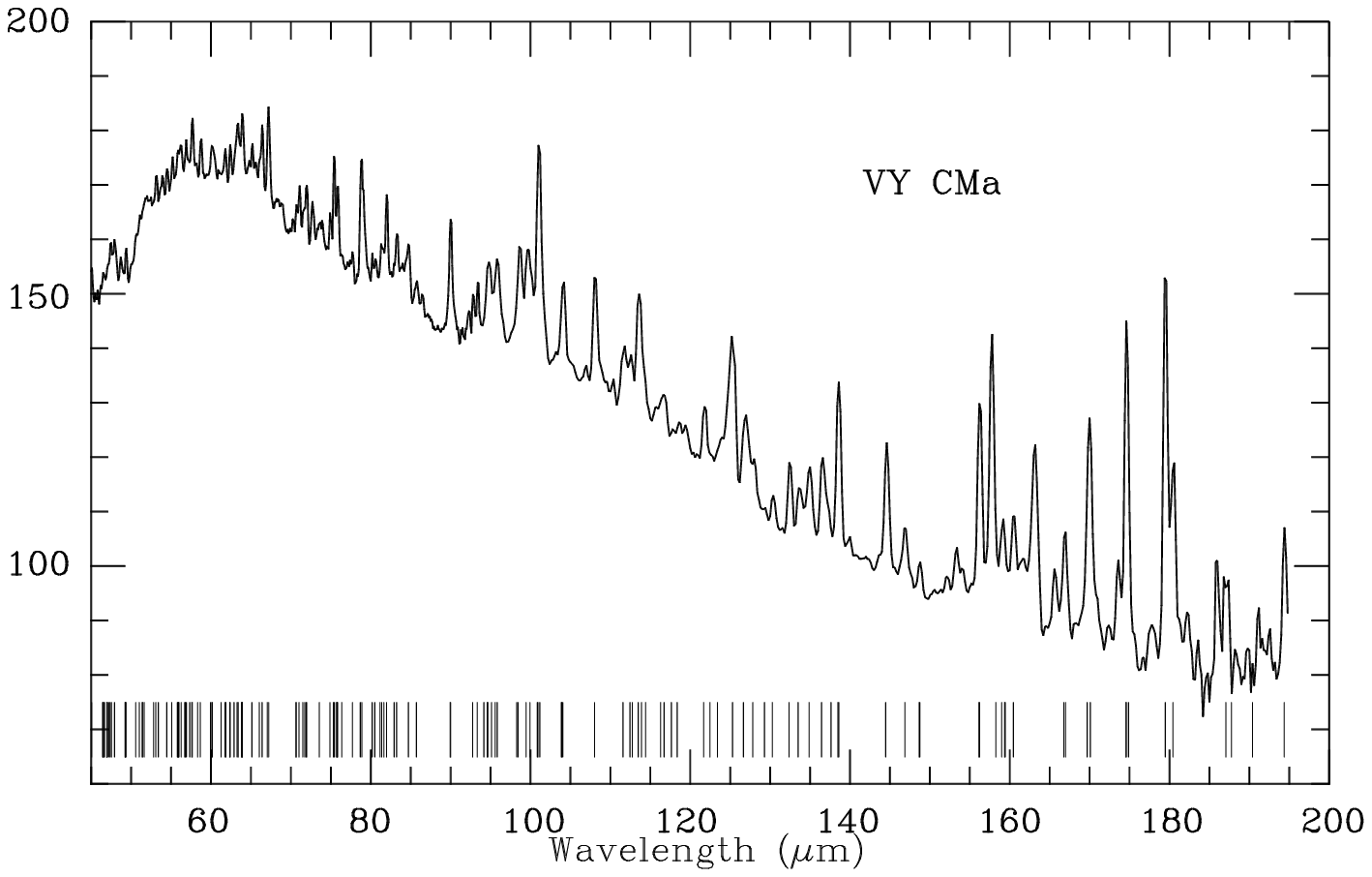}
\includegraphics[width=0.67\textwidth, angle=270]{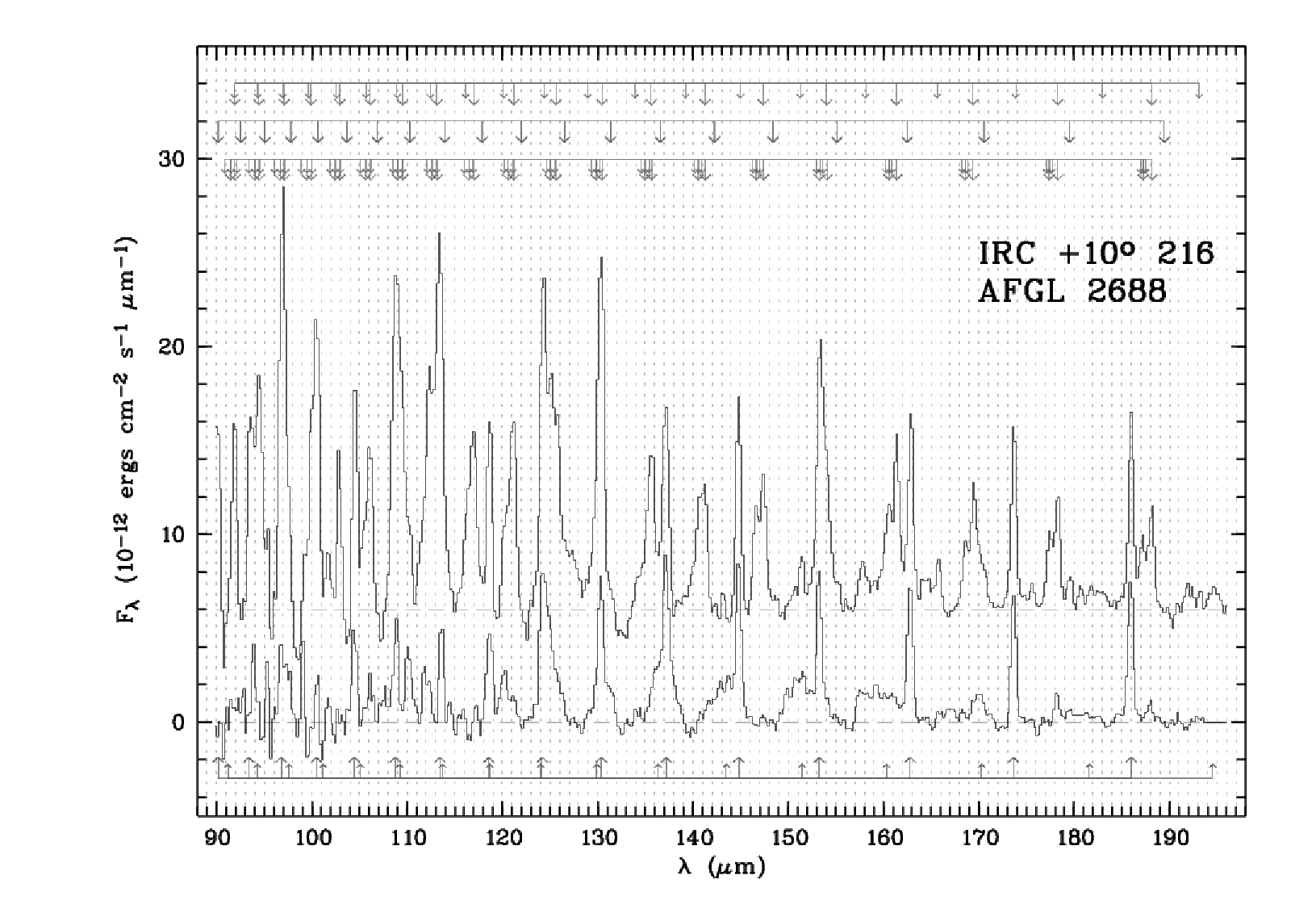}
\begin{center}
\caption[]{{\it Top:} The 43-197~$\mu$m {\em ISO} LWS spectrum of the M 
supergiant VY~CMa.
The observed flux (F$_{\lambda}$) has been multiplied by $\lambda^4$
to show this Rayleigh-Jeans region of the spectrum more clearly.
The tick marks at the bottom indicate the wavelengths of some of
the ortho- and para-H$_2$O rotational lines in this spectral region. 
Over 100 water 
lines are detected. {\it Bottom:} The continuum-subtracted 90-197~$\mu$m
{\em ISO} LWS spectra of the carbon star IRC+10$^{\rm o}$216 (upper
spectrum)
and of the C-rich post-AGB object AFGL~2688 (lower spectrum). The upward
pointing arrows indicate the positions of v=0
and v=1 rotational lines of CO. The downward pointing arrows
indicate the positions of rotational lines of HCN and H$^{13}$CN, as well 
as of vibrationally excited HCN rotational lines. See Cernicharo et al. 
(1996) and Cox et al. (1996) for more details on the carbon-rich source 
spectra.
}
\end{center}
\end{figure}

The {\em Herschel} SPIRE FTS will obtain spectra at up to R~=~1000 for a
large number of O-rich and C-rich sources from 200-650~$\mu$m, searching
particularly for new dust features that may be present. These features may
also occur in the spectra of star forming regions and galaxies, but the
best place to isolate and identify them is in the spectra of objects with
known chemistries, around which they have formed. In addition, the
continuum spectral properties of different dust species, particularly
their emissivity laws, have yet to be fully characterised in this spectral
region.

\subsection{The `missing mass' problem for intermediate mass stars}

An `average' planetary nebula has a central star mass of $\sim$0.6~M$_\odot$ 
and a nebular mass of $\sim$0.3~M$_\odot$. Population modelling predicts a
typical main sequence progenitor mass of 1.3~M$_\odot$, so about
0.4~M$_\odot$ appears to have been lost during earlier stages of
evolution. A much greater discrepancy exists for intermediate mass
progenitors. Populations of white dwarfs have been found in open clusters
which have main sequence turn-off masses of 6-8~M$_\odot$ (e.g. in
NGC~2516; see Weidemann 2000). So, $5-7$~M$_\odot$ must have been lost in
order to allow such stars to get below the Chandrasekhar limit, yet the
most massive PNe (e.g. NGC~6302, NGC~7027) contain no more than
2~M$_\odot$ of nebular material. So when was the rest of the mass lost
(and how do stars that develop degenerate cores know about the 
Chandrasekhar limit?) There is an obvious need for a comprehensive survey 
of the mass loss histories of evolved stars.

\begin{figure}
\begin{center}
\includegraphics[width=0.48\textwidth]{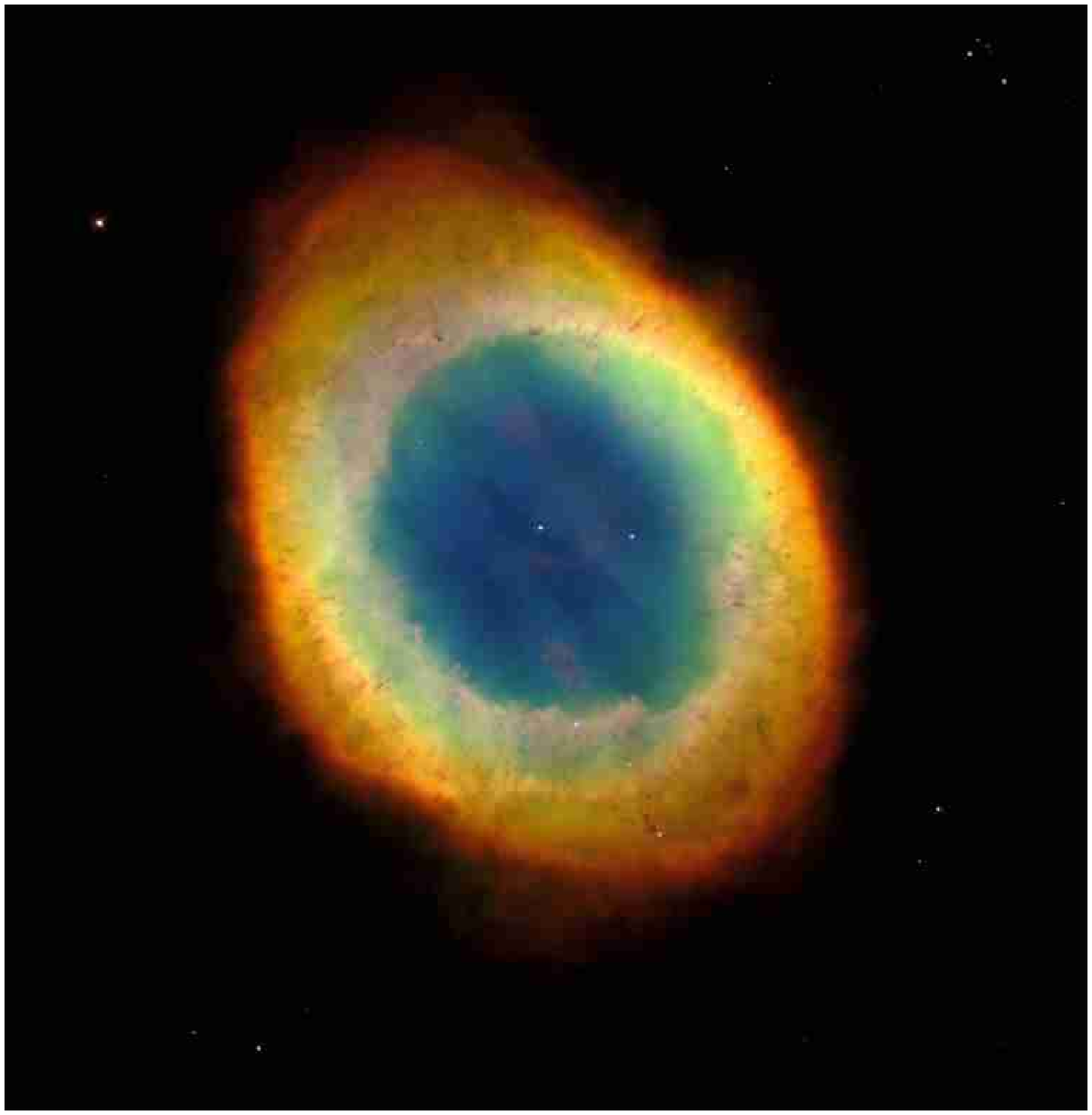}
\includegraphics[width=0.485\textwidth]{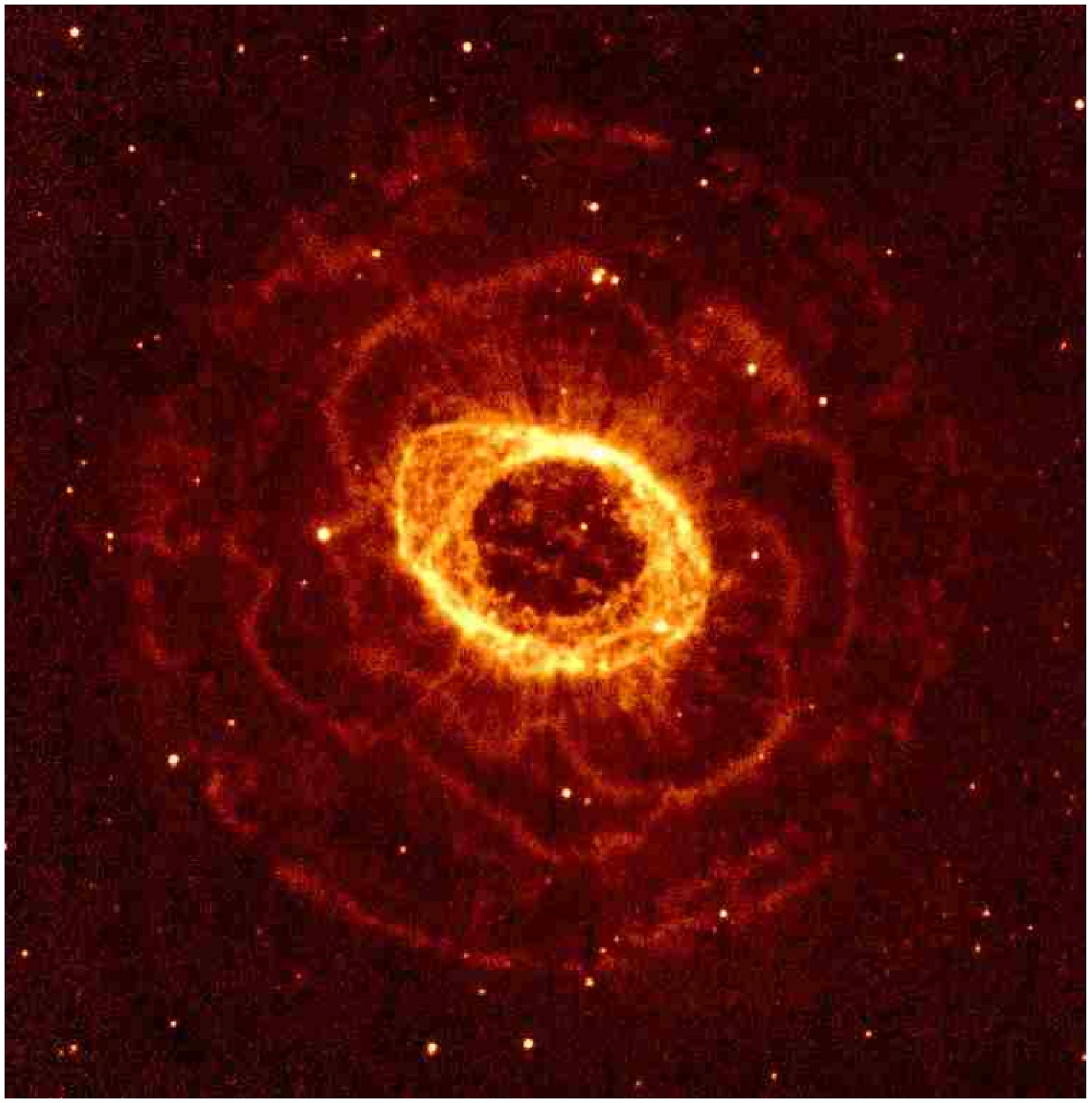}
\end{center}

\caption[]{Left: An {\em HST} WFPC2 publicity image of the Ring Nebula, 
NGC~6720, taken by combining exposures in each of three filters.
Red: F658N ([N~{\sc ii}]), green: F501N ([O~{\sc iii}]), Blue: F469N 
(He~{\sc ii}). The FoV is 2$\times$2~arcmin.
Right: a publicity image of the Ring Nebula seen observed in the 
H$_2$ v=1-0 S(1) 2.122-$\mu$m line with the MPA Omega near-IR
camera on the Calar~Alto 3.5-m telescope. The FoV
is 5$\times$5~arcmin. The bright inner emission corresponds to the 
optical nebula; the outer rosette-shaped filaments correspond to H$_2$
emission from material that is situated well beyond the optically emitting 
nebula. 
}
\end{figure}

The most sensitive method to search for ejected material around evolved
stars is to image the FIR/submm emission from dust particles in the ejecta
that are being heated by the interstellar radiation field to temperatures
of 20-30~K, peaking at far-IR wavelengths. Ionized gas can only be
detected following the onset of the planetary nebula phase, while cool
rarefied atomic or molecular gas can be extremely difficult to detect
except in favorable circumstances (Figure~11 shows the H$_2$-emitting halo
detected around the Ring Nebula). On the other hand, spatially extended
dust emission from ejected material is much easier to detect, during any
phase of evolution. Very extended dust shells have been detected around a
number of AGB stars, in {\em IRAS} and {\em ISO} far-infrared images (see
e.g. Izumiura et al. 1996, and references therein).

Compared to earlier facilities, the {\em Herschel Space Observatory} will
have a greatly enhanced sensitivity to extended emission at far-infrared
and submillimeter wavelengths. Following its launch in 2008/9,
one of its Guaranteed Time programmes will carry out mapping observations
aimed at detecting and determining the masses of extended dust shells
around a wide range of evolved star classes, in order to trace their mass
loss histories. Shells produced by past mass loss events over periods of
up to 40,000 years are potentially detectable and can yield information on
the mass loss process itself, e.g. whether it has been continuous or
episodic. Multi-wavelength photometric imaging can yield fluxes, dust
temperatures and dust shell masses. The PACS and SPIRE instruments will
obtain scanned maps of up to 30$\times$30 arcmin at 70, 110, 250, 350 and
520~$\mu$m for a large sample of targets, including AGB stars (O-rich and
C-rich), post-AGB objects and PNe. High galactic latitude targets will be
favoured, to minimise background confusion. High spectral resolution
follow-up observations of the [C~{\sc ii}] 158-$\mu$m line with {\em
Herschel}-HIFI can provide kinematic information on extended circumstellar
shells detected via imaging. Overall, these observations can lead to a
better understanding and quantification of the mass loss histories of low-
and intermediate-mass stars.

\section{Dust production by massive stars}

Where did the large quantities of dust detected in many high redshift
galaxies originate from? Bertoldi et al. (2003) detected redshifted warm
dust emission at millimeter wavelengths from three QSOs with z$>$6, i.e.
dust had formed less than 1~Gyr after the Big Bang. Dwek, Gallianio \&
Jones (2008) considered the case of the ultraluminous galaxy
SDSS~J1148+5251, at z~=~6.4. Its IR luminosity and dust mass were
estimated to be $2\times10^{13}$~L$_\odot$ and $2\times10^{8}$~M$_\odot$,
respectively, with its luminosity implying a current star formation rate
of $\sim$ 3000~M$_\odot$~yr$^{-1}$. At z~=~6.4, the Universe was only
900~Myr old; if the galaxy formed at z~=~10, then it is only 400~Myr old.
In fact, given its estimated dynamical mass of $5\times10^{10}$~M$_\odot$,
its current star formation rate would give an age of only 20~Myr. An age
of $20-400$~Myr would be insufficient for AGB stars to appear - only
massive stars would have had sufficient time to evolve and produce dust.
Elvis, Marengo \& Karovska (2002) suggested that QSO winds could reach
temperatures and pressures similar to those found around cool dust-forming
stars and that up to 10$^7$~M$_{\odot}$ of dust could be formed.
Markwick-Kemper et al. (2008) observed the z~=~0.466 broad absorption line
QSO PG~2112+059 with the {\em Spitzer} IRS and detected mid-IR emission
features which they attributed to amorphous and crystalline formed in the
quasar wind. It is not yet clear whether all dust-emitting high-z galaxies
possess such AGN central engines. If massive stars should turn out to be
the dominant sources of dust in high-z galaxies, which of their
evolutionary phases is the dominant dust producer? Is it the late-type
supergiant phase, the Luminous Blue Variable phase, the Wolf-Rayet phase,
or the final core-collapse supernova event?

\subsection{Late-type Supergiants and Hypergiants}

The M2~Iab supergiant $\alpha$ Orionis, with a luminosity of
$\sim2.5\times10^5$~L$_{\odot}$, has been estimated to have a gas mass
loss rate of $\sim 1.5\times10^{-5}$~M$_{\odot}$~yr$^{-1}$ (Jura \& Morris
1981). More luminous late type supergiants (sometimes dubbed
`hypergiants') can have even higher mass loss rates and are often
self-obscured by their own circumstellar dust at optical wavelengths,
e.g. the red hypergiants VY~CMa (Fig.~10), VX~Sgr and NML~Cyg. Similarly
high mass loss rates can be exhibited by yellow hypergiants, e.g.
$\rho$~Cas, IRC+10~420 and HR~8752. Such objects could potentially make a
very significant contribution to the dust enrichment of galaxies but
currently we do not know the duration of the yellow/red hypergiant phase,
nor the total population of such objects in our galaxy. Better statistics
from current optical and near-IR surveys, together with distances from
Gaia and more precise mass loss rate determinations using
improved wind modelling techniques, should lead to a much improved
understanding of the contribution of these objects to the dust and gas
evolution of galaxies.

\subsection{Luminous Blue Variables and Wolf-Rayet stars}

The ejecta nebulae around Luminous Blue Variables (LBVs) can contain large
masses of dust, as in the cases of $\eta$~Car and AG~Car (though not in
the case of P~Cygni); the 1840's outburst of $\eta$~Car has been estimated
to have produced 0.2~M$_\odot$ of dust (Morris et al. 1999). The M~1-67
ejecta nebula around the WN8 Wolf-Rayet star WR124 (Fig.~12) also contains
large quantities of dust and is thought to have originated from the
outburst of an LBV precursor.

\begin{figure}
\begin{center}
\includegraphics[width=0.75\textwidth]{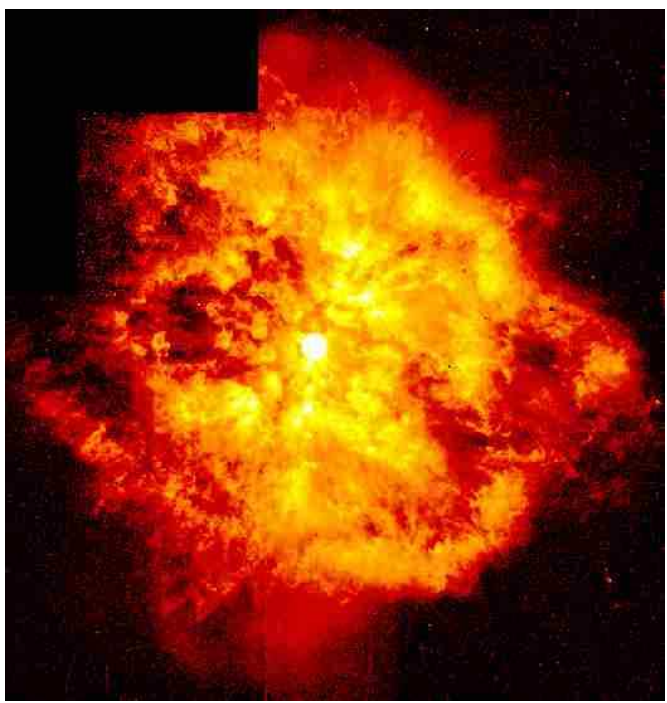}
\end{center}

\caption[]{{\em HST} WFPC2 F656N image of the nebula M~1-67 around
the WN8 Wolf-Rayet star WR124 (=BAC~209). For further details, see 
Grosdidier et al. (1998).} 
\end{figure}

The dust in most LBV nebulae appears to be dominated by oxygen-rich
silicate grains. Massive carbon-rich WC9 Wolf-Rayet stars often show hot
($\sim$900~K) featureless dust emission that has been attributed to carbon
grains. How does dust form in outflows from stars whose effective
temperatures are in the region of 30,000~K? The answer was found by
Tuthill et al. whose masked-aperture Keck imaging at 2.27-$\mu$m revealed
a rotating pinwheel plume of dust emission in the WC9+OB binary system
WR104 (Fig~13). The dust appears to be formed in the compressed shock
interaction region (hotspot) between the stellar winds of the WC9 primary
and the OB secondary, the relative motions of the two stars creating an
Archimedean spiral. Similar pinwheel structures have since been found
around several more WC9 systems, including two located in the Galactic
Center Quintuplet Cluster (Tuthill et al. 2006). The overall contribution
of late WC-type Wolf-Rayet stars to the dust enrichment of galaxies is
currently extremely uncertain. Future submm observations by ALMA can help
to quantify their overall dust production rates, plus search for molecular
emission from the hot-spot shocked wind compression region, while the GAIA
parallax survey of the Galaxy should enable their total numbers to be
more accurately estimated.

\begin{figure}
\begin{center}
\includegraphics[width=0.6\textwidth]{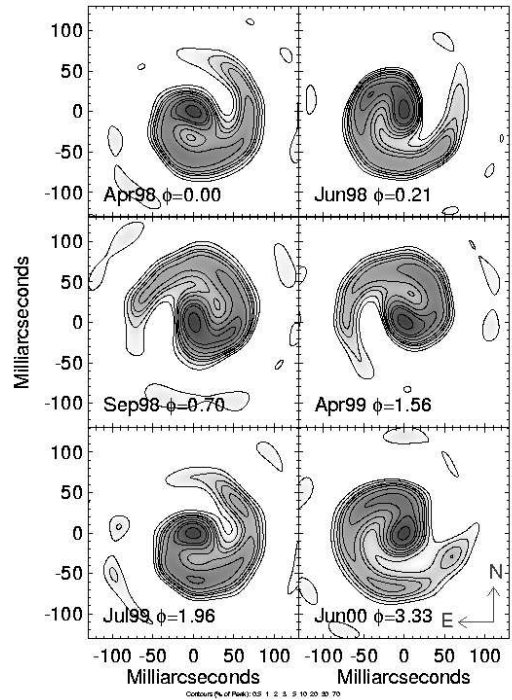}
\end{center}

\caption[]{Keck aperture-mask 2.27-$\mu$m imaging of a rotating pinwheel 
plume of dust emission from the WC9+OB binary WR~104. The dust is formed 
in the compressed shock interaction region (hotspot) between the two stellar 
winds. The relative motions of the two stars creates an Archimedean 
spiral. From Tuthill, Monnier \& Danchi (2002).} 
\end{figure}

\subsection{Core collapse supernovae}

There is plenty of evidence that supernovae can synthesise dust particles 
that are able to survive the shock buffeting that must take place 
following their formation. The isotopic analysis of pre-solar meteoritic 
grain inclusions (those that have non-solar ratios) has found many 
examples that are dominated by r-process isotopes, indicating a supernova 
origin (e.g. Clayton, Amari \& Zinner 1997). The onion-skin abundance 
structure of a pre-supernova massive star means that different layers of 
the ejecta can have C/O$<$1 and C/O$>$1, allowing the 
formation of O-rich grains and C-rich grains in the respective zones. 
Infrared photometry and spectrophotometry of SN~1987A demonstrated the 
onset of thermal dust emission by day 615 (Bouchet \& Danziger 1993; 
Wooden et al. 1993), as shown in Figure~14, the emission being attributed 
to newly formed dust particles in the ejecta.

\begin{figure}
\begin{center}
\includegraphics[width=0.85\textwidth]{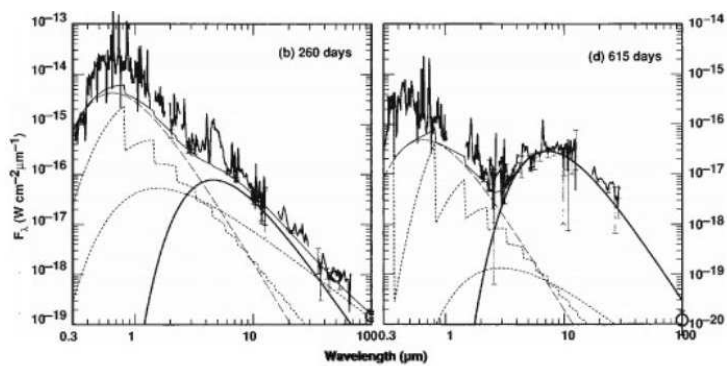}
\end{center}

\caption[]{Optical and KAO IR spectrophotometry of SN~1987A, illustrating 
the definite onset of  thermal dust emission by day 615. From Wooden et 
al. (1993), who derived a lower limit of 10$^{-4}$~M$_{\odot}$ to the 
mass of dust that had formed by day 775.}
\end{figure}

From dust nucleation modelling, Todini \& Ferrara (2001) predicted that 
0.08-1.0~M$_\odot$ of dust could condense in the ejecta of of a typical 
high-redshift core collapse supernova within a few years of outburst, 
corresponding to a condensation efficiency for the available refractory 
elements of $>0.2$. Similarly high condensation efficiencies appear to be 
required to explain the $\sim10^8$ solar masses of dust deduced to exist 
in high redshift QSOs (Morgan \& Edmunds 2003; Dwek et al. 2008). 
However, prior to the launch of {\em Spitzer}, for the handful of recent 
core-collapse SNe for which dust formation had been inferred, the derived 
masses of newly formed dust were $\leq10^{-3}$~M$_\odot$ (e.g. the 
examples shown in Figs. 14 and 15)

A different approach to estimating how much dust can be formed in the 
ejecta of a core-collapse SN was taken Dunne et al. (2003) and Morgan et 
al. (2003), who used SCUBA submillimeter maps to deduce that 
$1-2$~M$_\odot$ of dust were present in the Cas~A and Kepler supernova 
remnants (SNRs), both of which were less than 400 years old. However, 
Krause et al. (2004) argued that most of the observed submm emission 
observed from Cas~A originated from a foreground molecular cloud that 
could be seen in CO maps. Deeper observations of more young SNRs across a 
wide wavelength range are planned in {\em Herschel Space Observatory} 
Guaranteed Time; $57-650~\mu$m PACS and SPIRE photometric 
and spectroscopic maps will be obtained of five galactic SNRs having ages 
of less than 1000~yrs (Cas A, Kepler, Tycho, Crab and 3C58)

Returning to observations of the dust formation phase itself (t$<$1000 
days), there are currently three methods for inferring the formation of 
dust in supernova ejecta:

\begin{figure}
\begin{center}
\includegraphics[width=0.45\textwidth]{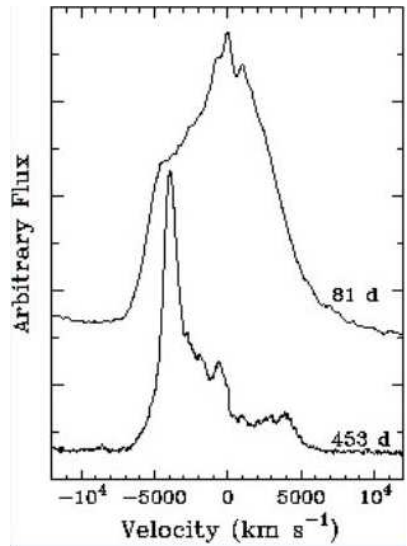}
\end{center}

\caption[]{The H$\alpha$ line profile of SN~1998S at days 81 and 453 
(Leonard et al. 2000), 
illustrating the loss of flux on the red side of the profile, attributed 
to absorption by newly formed dust preferentially removing photons from 
the far side of the ejecta. Pozzo et al. (2004) estimated that 
10$^{-3}$~M$_\odot$ of dust had formed.}
\end{figure}

\begin{enumerate}
\item via the detection of thermal IR emission from the newly formed dust.  
However, the use of this method alone can be compromised by 
pre-existing nearby dust (e.g. circumstellar dust), which can be heated 
by the supernova light flash.

\item via the detection of a dip in the SN light curve that can be 
attributed to extinction by newly formed dust. Pre-existing dust cannot 
produce such a dip.

\item via the detection of the development of a red-blue asymmetry in the 
SN emission line profiles, attributable to the removal by newly formed 
dust of some of the redshifted emission from the far side of the SN ejecta 
(Lucy et al. 1989; see Fig.~15).
\end{enumerate}
       
Method (1) is normally required if dust masses are to be quantified
but ideally it ought to be supported by one or both of (2) and (3).

Since the launch of the {\em Spitzer Space Telescope}, two teams have been
conducting observing programmes to study the spectral evolution of young
supernovae. Examples of the spectra of SNe less than 250 days after
outburst are shown in Figure~16. The day~135 IRS spectrum of SN~2005df
(Gerardy et al. 2007) shows numerous ionic fine structure lines, including
lines from radioactive cobalt and nickel, whose decay is the main heating
source for the ejecta. The day~214 IRS spectrum of SN~2005af (Kotak et al.
2006) is dominated by the $v~=~1-0$ band-head of gaseous SiO and is very
similar to that of SN~1987A at a similar epoch.

\begin{figure}
\begin{center}
\includegraphics[width=0.48\textwidth]{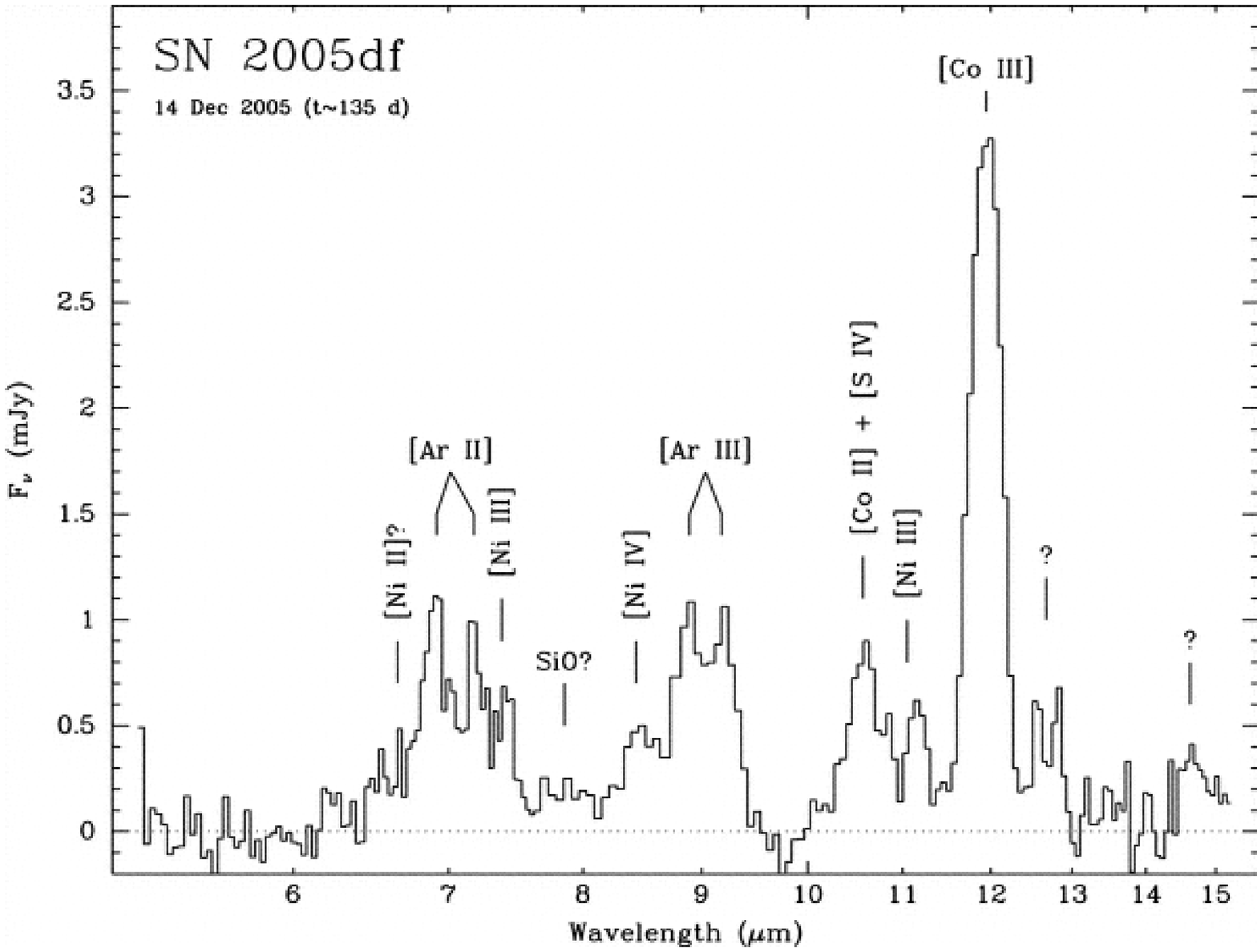}
\includegraphics[width=0.5\textwidth]{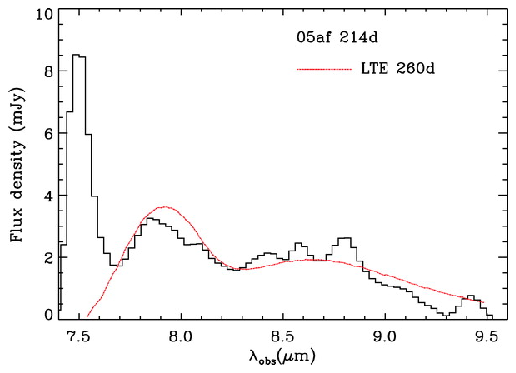}
\end{center}

\caption[]{Left: The {\em Spitzer} IRS spectrum of SN~2005df at day 135, 
showing ionic fine structure emission lines from radioactive cobalt and 
nickel (Gerardy et al. 2007). Right: The {\em Spitzer} IRS spectrum of 
SN~2005af 
at day 214, showing the 8-$\mu$m v=1-0 bandhead of gaseous SiO (Kotak et 
al. 2006). By analogy to SN~1987A, this was suggested to be a precursor 
to dust formation.}
\end{figure}

Early in the {\em Spitzer} mission, dust emission from the Type~IIP 
supernova SN~2002hh was detected in SINGS IRAC images and confirmed by 
higher angular resolution Gemini Michelle imaging (Barlow et al. 2005). 
The day~600 spectral energy distribution (SED) could be fitted by a 290~K 
blackbody (Fig.~17, upper), yielding a minimum emitting radius of R$_{\rm 
min}\sim10^{17}$~cm and and a luminosity of L$=1.6\times10^7$~L$_\odot$. 
A more realistic $\lambda^{-1}$ grain emissivity gave R$_{\rm min}$ = 
5$\times10^{17}$~cm. This was far too large for the emitting dust to have 
formed in the ejecta (it would have taken $>10$~yrs for material in the 
ejecta to reach this radius). It was therefore inferred that the emitting 
dust must have been pre-existing. All three of the model fits shown in 
Fig.~17 yielded total emitting dust masses in the range 
$0.10-0.15$~M$_\odot$. SN~2002hh is a Type~IIP (plateau) supernova, whose 
very extended optical light curve (Welch et al. 2007) appears explicable 
in terms of a just-resolved light echo that has been revealed by {\em HST} 
ACS/HRC images (Sugerman 2005). Preliminary analysis indicates that the 
echo has occurred from a thick dust distribution that is located about 2-8 
light years ($2-8\times10^{18}$~cm) in front of the supernova.

\begin{figure}
\begin{center}
\includegraphics[width=0.7\textwidth]{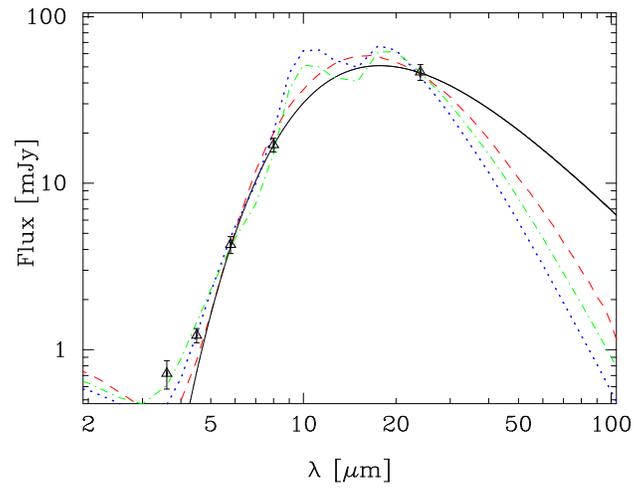}
\end{center}

\caption[]{The measured day-600 {\em Spitzer} $3.6-24~\mu$m 
fluxes for SN~2002hh, in NGC~6946, are
shown as open triangles, with vertical bars indicating the flux
uncertainties. The solid black line is a 290-K blackbody normalised to the
8.0-$\mu$m flux point. The dashed, dotted and dash-dotted lines correspond
to radiative transfer models with differing amounts of silicates and 
amorphous carbon; see Barlow et al. (2005).
}
\end{figure}

\begin{figure}
\begin{center}
\includegraphics[width=0.95\textwidth]{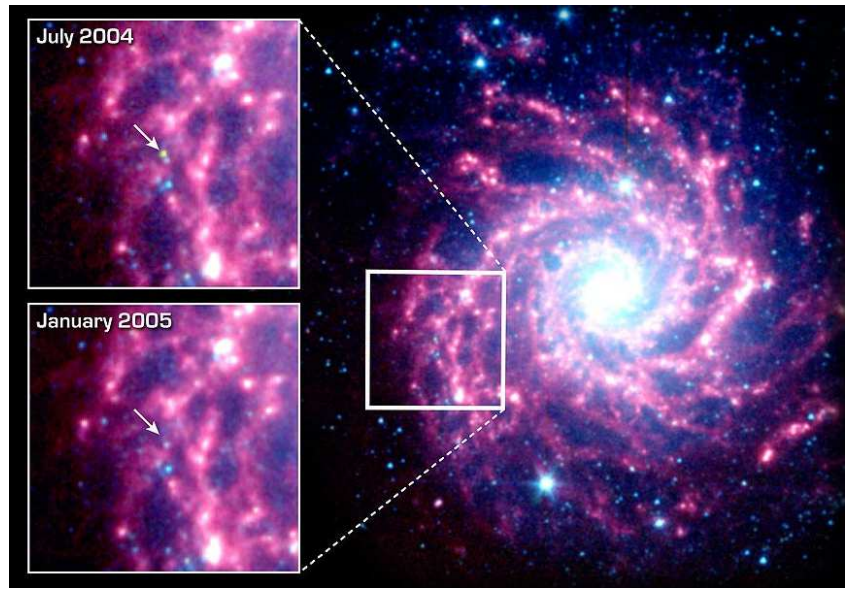}
\end{center}

\caption[]{A {\em Spitzer} SINGS multi-band IRAC image of NGC~628.
There is a clear detection of SN~2003gd on day-499 (upper inset)
relative to day-670 (lower inset). See Sugerman et al. (2006).
}
\end{figure}

Sugerman et al. (2006) detected the onset of dust emission from SN~2003gd 
in NGC~628 (Messier 74; see Fig.~18); unlike SN~2002hh, the emitting dust 
was inferred 
to have formed inside the supernova ejecta. Between days 157 and 493 the 
H$\alpha$ feature developed an asymmetric profile, with a reduction in 
flux on the 
red side. This was attributed to dust forming in the ejecta preferentially 
extinguishing emission from receding (red-shifted) gas. There was 
also an increase in optical extinction after day~500, as evidenced by a 
dip in its light curve from that date, similar to the behaviour of 
SN~1987A at the same epochs. Additional extinction by dust was inferred to 
have occurred after day 500 for both SNe, and for SN~2003gd corresponded 
to $0.25-0.5$ magnitudes in the R-band on day~500, and to $0.8-1.9$ 
magnitudes on day 678.

Both smooth and clumped SN ejecta model fits to the SN~2003gd 
observations were presented by Sugerman et al. (2006), using a 3D Monte 
Carlo radiative transfer code with a mother-grid of 61$^3$ cells (the 
mother cells that contained clumps were resolved by a subgrid of 5$^3$ 
cells). To match both the optical-IR SEDs and the derived R-band 
extinction estimates for the SN ejecta, their day~499 data could be fitted 
with smoothly distributed dust having a dust mass of 
$2\times10^{-4}$~M$_\odot$, whereas up to $2\times10^{-3}$~M$_\odot$ of 
dust could be accommodated by a clumped dust model. For day~678, their
best-fit smooth dust model required $3\times10^{-3}$~M$_\odot$ of dust, 
while up to $2\times10^{-2}$~M$_\odot$ could be accommodated by a clumpy 
model, the latter implying a heavy element condensation efficiency of 
about 10\%. The {\em Spitzer} observations of SN~2003gd have also been 
studied by Meikle et al. (2007).

\begin{table}
\centering
\caption{Thermal infrared studies of core collapse supernovae -- results 
so far}
\begin{tabular}{llll}

\hline
Name & D(Mpc) & Progenitor mass, M$_\odot$ & Dust emission? \\
\hline
SN~1987A   & 0.05  & 16-22 & Yes \\
SN~1999bw  & 14.5  & unknown & Yes, but very late \\
SN~2002hh  & 5.6   & 8-14  & Yes, but pre-existing \\
SN~2003gd  & 9.3   & 6-12  & Yes \\
SN~2004dj  & 3.3   & 12-15 & Maybe; pre-existing? \\
SN~2004et  & 5.6   & 13-20 & Yes \\
SN~2005cs  & 8.0   & 7-12  & No  \\
\hline
\end{tabular}
\end{table}

Table~1 summarises the results to date of mid-infrared searches for dust 
emission from young supernovae. Apart from SN~1987A, all are based on {\em 
Spitzer} observations, supplemented in two cases by Gemini North Michelle 
observations. Although {\em Spitzer} has delivered a very large increase 
in mid-infrared sensitivity relative to prior facilities, its distance 
limit for the detection of dust forming around young SNe is 
effectively 10-15~Mpc, corresponding to a volume within which relatively 
few new SNe occur each year. All of the SNe listed in Table~1 are of
Type~II, from progenitors with masses $<20$~M$_\odot$. No Type~Ib or Ic 
SNe, whose immediate precursors are believed to be H-deficient Wolf-Rayet 
stars descended from much more massive stars, have so far been close 
enough to be detected by {\em Spitzer}, although the Type~Ib SN~2006jc
has shown evidence for dust formation via the development of red-blue
emission line asymmetries and a transient far-red and near-IR continuum
excess, interpreted by Smith, Foley \& Filippenko (2008)
as due to dust formation in the dense region created by the impact between
the SN ejecta and slower moving material from an LBV-type eruption that 
was discovered two years before the supernova outburst.

For point source imaging, {\em Spitzer}'s IRAC is 135$\times$ more 
sensitive at 8~$\mu$m than mid-IR instruments on ground-based 8-m 
telescopes (due to {\em Spitzer}'s vastly lower thermal backgrounds). For 
point source imaging, {\em JWST}-MIRI is expected to be $\sim40\times$ 
more sensitive at 8~$\mu$m than IRAC
(see www.stsci.edu/jwst/science/sensitivity/). In addition to these 
sensitivity gains, the 8$\times$ higher angular resolution of {\em 
JWST}-MIRI compared to {\em Spitzer}-IRAC will greatly reduce point source 
confusion effects in dense starfields, such as encountered when observing 
galaxies. 

\begin{figure}
\begin{center}
\includegraphics[width=0.50\textwidth]{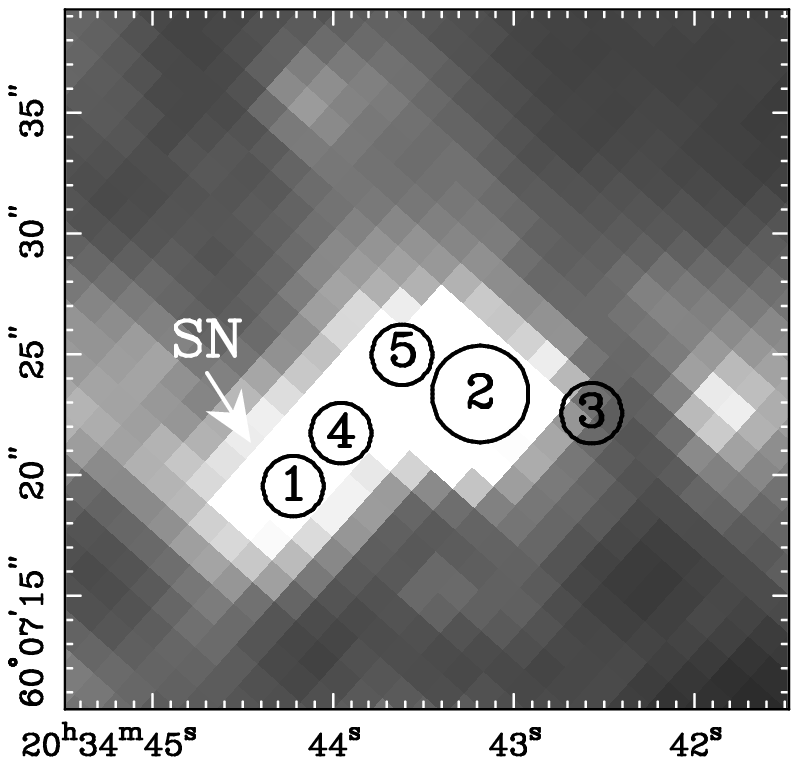}
\includegraphics[width=0.48\textwidth]{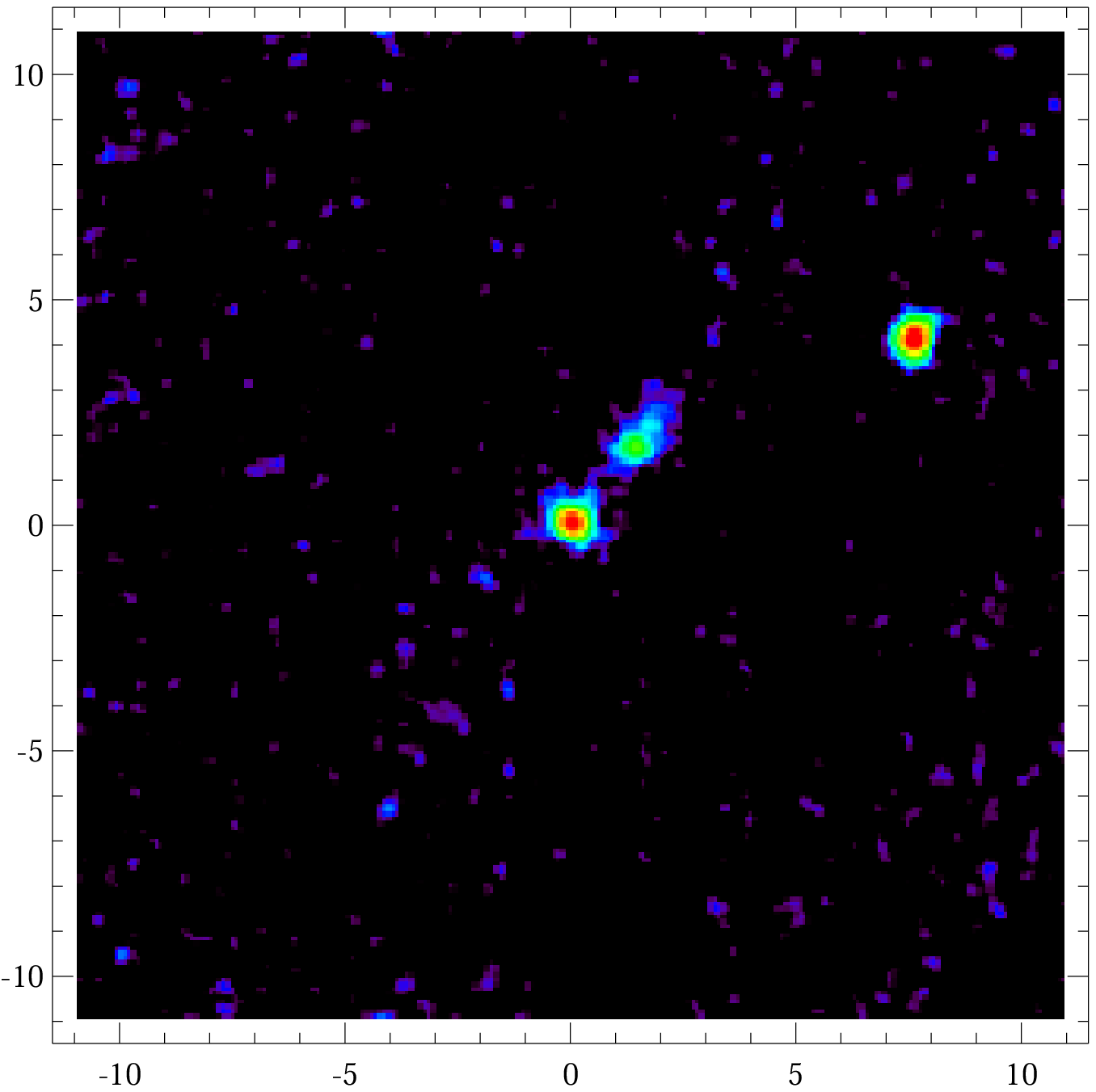}
\end{center}

\caption[]{Left: Day-590 {\em Spitzer} IRAC 8-$\mu$m image of the 
region around SN~2002hh. With {\em Spitzer's} 2.4~arsec angular 
resolution, five 
different sources are blended together. Right: the 0.3~arsec angular
resolution of this day-698 Gemini-N Michelle 11-$\mu$m image
completely resolves the sources (the SN is at the center), illustrating 
how in the case of crowded-field galaxy observations the high angular 
resolution of JWST-MIRI will supplement its raw sensitivity.
}
\end{figure}

The angular resolution advantages of a $>$6-m class telescope 
compared to a 0.85-m telescope are illustrated in Fig.~19, where an IRAC 
8-$\mu$m image of the field of SN~2002h, in NGC~6946, is compared to a 
8$\times$ higher resolution Gemini Michelle 11-$\mu$m image of the same 
field. The five sources that are blended together in the IRAC image are 
completely resolved from each other in the Gemini image. MIRI's much 
greater sensitivity and angular resolution should enable SNe out to nearly 
200~Mpc to be detected at mid-IR wavelengths, corresponding to a volume 
$\sim$1000 times larger than for {\em Spitzer}. So MIRI will be able to 
quickly observe large numbers of new SNe of all classes, both 
photometrically and spectroscopically, enabling the dust contribution
by each class to be accurately assessed. \\

\noindent
I would like to thank Christoffel Waelkens and Xander Tielens for
their comments and suggestions.

%

\end{document}